\documentclass[a4paper,twoside]{article}

\usepackage{epsfig}
\usepackage{subfigure}
\usepackage{calc}
\usepackage{amssymb}
\usepackage{amstext}
\usepackage{amsmath}
\usepackage{amsthm}
\usepackage{multicol}
\usepackage{pslatex}
\usepackage{apalike}
\usepackage{SCITEPRESS}
\usepackage[small]{caption}
\usepackage{flexisym}
\usepackage{pdfcomment}

\usepackage{algorithm}
\usepackage{algpseudocode}

\usepackage{tabularx}

\DeclareMathOperator*{\argmin}{arg\,min}

\algnewcommand{\algorithmicgoto}{\textbf{go to}}%
\algnewcommand{\Goto}[1]{\algorithmicgoto~\ref{#1}}%

\begin{document}

\title{Executing Bag of Distributed Tasks on Virtually Unlimited Cloud Resources}

\author{\authorname{Long Thai, Blesson Varghese and Adam Barker}
\affiliation{School of Computer Science, University of St Andrews, Fife, UK}
\email{\{ltt2, varghese, adam.barker\}@st-andrews.ac.uk}
}


\abstract{Bag-of-Distributed-Tasks (BoDT) application is the collection of identical and independent tasks each of which requires a piece of input data located around the world. As a result, Cloud computing offers an effective way to execute BoT application as it not only consists of multiple geographically distributed data centres but also allows a user to pay for what she actually uses only. In this paper, BoDT on the Cloud using virtually unlimited cloud resources. A heuristic algorithm is proposed to find an execution plan that takes budget constraints into account. Compared with other approaches, with the same given budget, our algorithm is able to reduce the overall execution time up to 50\%.
}

\onecolumn \maketitle \normalsize \vfill

\section{\uppercase{Introduction}}
\label{sec:introduction}

Bag-of-Tasks (BoT) is the collection of identical and independent tasks. In other works, tasks of a BoT application can be executed by the same application but in any order. Bag-of-Distributed-Tasks (BoDT) is a subset of BoT in which each task requires data from somewhere around the globe. The location where a task is executed is essential for keeping the execution time of the BoDT low, since data is transferred from a geographically distributed location. It is ideal to assign tasks to locations that would be in geographically close proximity to the data.


The centralised approach for executing BoDT, in which data from multiple locations are transferred and executed at a single location, tends to be ineffective since some data resides very far from the selected location and takes a long time to be downloaded. Another approach is to group the tasks of the BoDT in such a way that each group can be executed near the location of the data. However, this approach requires an infrastructure which is decentralised and globally distributed. Cloud computing is ideal suited for this since public cloud providers have multiple data centres which are globally distributed. Furthermore, due to its pay-as-you-go scheme, using Cloud computing is cost effective as a user only pays for Virtual Machines (VMs) that are required.

Cloud computing can facilitate the execution of BoDT, and at the same time introduce the challenge of assigning tasks to VMs by considering the location for processing each task, the user's budget constraint, as well as the desired performance, i.e. execution time, for executing the task. In an ideal case, it is expected that maximum performance is obtained while minimising the costs.

In our previous paper \cite{Thai:2014:CloudCom}, we approached this problem by assuming limited resources were available. However, as Cloud provider offers virtually unlimited resources, the limit should be determined based on the user's budget constraint. In this paper, we present our approach for executing BoDT on the Cloud on virtually unlimited resources and is only limited by a user specified budget constraint. Compared with other approaches, with the same given budget, our algorithm is able to reduce the overall execution time up to 50\%.

The contributions of this paper are i) the complete mathematical model of executing a BoDT application on the Cloud with budget constraint, ii) the heuristic algorithm which assigns tasks to Cloud resources based on their geographical locations, and iii) the evaluation comparing our approach with the centralised and the round robin approaches.

The remainder of paper is structured as follow. Section II presents the mathematical model of the problem. Section III introduces the heuristic algorithms producing an execution plan based on the user's budget constraint. Section IV evaluates the approach. Section V presents the related work. Finally, this paper is concluded in section VI.

\section{\uppercase{Problem Modelling}}
Let $L = \{l_1 ... l_m\}$ be the list of Cloud locations, i.e. location of Cloud provider's data centres, and $VM = \{vm_1 ...\}$ be the list of Cloud VMs. For $vm \in VM$, $l_{vm} \in L$ denotes the location in which $vm$ is deployed. Let $VM_l \subset VM$ be the list of all VMs deployed at location $l \in L$. The number of items in $VM$ is not fixed since a user can initiate as many VMs as possible.

Let $T = \{t_1 ... t_n\}$ be the list of tasks, and $size_t$ denote the size of a task. The time (in seconds) taken to transfer data from a task's location to a Cloud location is denoted as $trans_{t, l}$. Similarly, $trans_{t, vm}$ for $vm \in VM$ is the cost of moving $t$ to $vm$ (or to a location on which $vm$ in running; $trans_{t, vm} = trans_{t, l_{vm}}$). We assume that there is only one type of VM is used, hence, the cost of processing one unit of data is identical and is denoted as $comp$.

The time taken to execute task $t$ at $vm$ is:
\begin{equation}
	exec_{t, vm} = exec_{t, l_{vm}} = (trans_{t, vm} + comp) \times size_t
\end{equation}

Let $T_{vm} \subset T$ be the list of tasks executed in $vm \in VM$. All tasks must be executed and is represented as the following constraint:
\begin{equation} \label{completeness}
	\bigcup_{vm \in VM}{T_{vm}} = T
\end{equation}

One task should not be executed in more than one location expressed as an additional constraint:
\begin{equation} \label{no_duplicate}
	T_i \cap T_j = \emptyset \quad \text{for $i, j \in VM$ and $i \neq j$}
\end{equation}

The execution time of all tasks on $vm \in VM$ is:
\begin{equation}
	exec_{T_{vm}} = \sum_{t \in T_{vm}}{exec_{t, vm}}
\end{equation}




As it takes some times to create a VM, the overhead associated with the start up of each VM denoted as $start\_up$. The execution time of $vm \in VM$ to execute all tasks in $T_{vm}$ is:
\begin{equation}\label{eq:exec_vm}
	exec_{vm} = start\_up + exec_{T_{vm}}
\end{equation}

It should be noted that Equation \ref{eq:exec_vm} can only be applied if there are task(s) assign to a VM, i.e. $T_{vm} \neq \emptyset$. Otherwise, it is unnecessary to create a VM, thus its execution time is zero.

Assuming each VM is charged by hour, i.e. 3600 seconds, the number of charged time blocks is:
\begin{equation}\label{eq:time_block}
	tb_{vm} = \lceil \frac{exec_{vm}}{3600} \rceil
\end{equation}

Equation \ref{eq:time_block} contains the ceiling function, which means the execution time is rounded up to the nearest hour in order to calculate the number of used time blocks. In other words, a user has to pay for a full hour even if only a fraction of the hour is used.

Let $P = \{T_{vm_1} ... T_{vm_p}\}$ be the execution plan, whose each item is a group of tasks assigned to one $vm \in VM$. Let $VM_P$ denote the list of VMs used by execution plan $P$. Similarly, let $L_P$ be the list of locations where all VMs of plan $P$ are deployed. Moreover, $P_l$ denotes the execution plan for location $l \in L$, which means $L_{P_l} = \{l\}$ and $VM_{P_l} = VM_l$.

As all VMs are running in parallel, the execution time of a plan is equal to slowest VM's:
\begin{equation}
	exec_P = \max_{vm \in VM_P}{exec_{vm}}
\end{equation}

The total number of time blocks used is the sum of the time blocks used by each VM, represented as:
\begin{equation}
	tb_P = \sum_{vm \in VM_P}{tb_{vm}}
\end{equation}

The budget constraint is the amount of money that a user is willing to pay for executing the BoDT. Even though Cloud providers charge users for using compute time on virtual machines and transferring data, only the renting cost is considered as the amount of downloaded is unchanged for any given problem, i.e. regardless the execution plan, the same amount of data is downloaded, thus the data transferring cost.

The budget constraint is mapped onto the number of allowed time blocks $tb_b$ by dividing the budget to the cost of one time block (this is possible, because of the assumption that there is only one VM type). Hence, the problem of maximising the performance of executing a BoDT on the Cloud with a given budget constraint is to find an execution plan $P$ in order to minimise $exec_P$ while keeping $tb_P = tb_b$ and satisfying constraints in Equations \ref{completeness} and \ref{no_duplicate}.

\section{\uppercase{Algorithms}}

As stated in the previous section, the optimal plan for executing BoDT on the Cloud with budget constraint can be found by solving the mathematical model. However, solving the mathematical model can take the considerable amount of time since it involves considering multiple possibilities of assigning tasks to different VMs at different Cloud locations. In this section, we propose an alternative approach as the heuristic algorithm for finding an executing plan for a BoDT based on a user's budget constraint.


\subsection{Select Initial Number of VMs at Each Location}
The main idea of the approach presented in this paper is to specify a set of VMs to each location, then to reduce them until the total number of VMs across all locations is $tb_b$.

In order to determine the initial number of VMs at each location, we make an assumption that it is possible to limit each VM to be executing in one time block, i.e. if a VM finishes its execution in more than one time block, its tasks can be split and scheduled onto two VMs. Then, the total number of time blocks is equal to the total number of VMs across all locations. Thus, the constraint $tb_b$ also limits the total number of VMs, each of which uses no more than one time block. Hence, initially, the number of VMs at each location, i.e. $\overline{VM_l}$ for $l \in L$, can be set to $tb_b$.

\subsection{Find Execution Plan based on Budget Constraint}
Let $P_{nl}$ be the plan in which tasks are assigned to their nearest location, i.e. the location in which $exec_{t, l}$ is minimum. Each item in $P_{nl}$ represents the list of tasks assigned to a location (not a VM).

\begin{algorithm}
	\caption{Find Execution Plan based on Budget Constraint}
	\label{al:find}
	\begin{algorithmic}[1]
		\Function{FIND\_PLAN}{$tb_b, P_{nl}, VM$}
			\State $P \gets \emptyset$
			\For{$l \in L_{P_{nl}}$}	\label{al:find:nearest:loop}
				\State $P_l \gets ASSIGN(T_l, VM_l)$
				\If{$tb_{P_l} > tb_b$}	\label{al:find:nearest:check}
					\State $FAIL$
				\EndIf
				\State $P \gets P_l$
			\EndFor	\label{al:find:nearest:end_loop}

			\State $P \gets REDUCE(P, \emptyset, TRUE)$			\label{al:find:reassign:start}
			\If{$tb_P > tb_b$}
				\State $P \gets REDUCE(P, \emptyset, FALSE)$
			\EndIf												\label{al:find:reassign:end}

			\If{$tb_P > tb_b$}	\label{al:find:reassign:check}
				\State $FAIL$	\label{al:find:reassign:fail}
			\EndIf

			\State $P \gets BALANCE(P)$ \label{al:find:balance}

			\State \Return $P$
		\EndFunction
	\end{algorithmic}
\end{algorithm}

Algorithm \ref{al:find} finds a plan with minimum execution time based on the budget constraint $tb_b$. The nearest plan $P_{nl}$ and the initial list of virtual machines $VM$ are provided as input. The algorithm uses three functions, namely $ASSIGN$, $REDUCE$ and $BALANCE$.

First of all, the algorithm assigns tasks to VMs deployed in their nearest locations (From Line \ref{al:find:nearest:loop} to \ref{al:find:nearest:end_loop}). Line \ref{al:find:nearest:check} checks if the number of used time block in a location is more than the budget constraint. If that is the case, then it is impossible to find an execution plan satisfying the given budget constraint.

Secondly, some VMs are removed by moving its tasks to other ones until the budget constraint is satisfied (From Line \ref{al:find:reassign:start} to \ref{al:find:reassign:end}). The reassignment can be performed between VMs in the same location or across multiple locations. If, after reducing, the number of VMs is still higher than $tb_b$, it is impossible to satisfy the budget constraint (Lines \ref{al:find:reassign:check} and \ref{al:find:reassign:fail}).

Finally, as the execution times between VMs are different (for example, one VM can take longer to finish than the other ones) it is necessary to balance out the execution times between all VMs so that they can finish at the same time, thus reduce the overall execution time (Line \ref{al:find:balance}).

\subsection{Assign Tasks to VMs}

\begin{algorithm}
	\caption{Assign Tasks to VMs}
	\label{al:assign}
	\begin{algorithmic}[1]
		\Function{ASSIGN}{$T', VM'$}
			\State $T' \gets T'$ sorted by $-exec_{t, l}$ for $t \in T'$						\label{al:assign:tasks_sort}
			\For{$t \in T'$}																	\label{al:assign:loop}
				\State $VM_0 \gets VM'$ filtered $exec_{vm} + exec_{t, vm} \leq 3600$			\label{al:assign:filter}
				\If{$VM_0 = \emptyset$}															\label{al:assign:check}
					\State $FAIL$																\label{al:assign:fail}
				\EndIf
				\State $VM_0 \gets VM'$ sorted by $(trans_{t, vm}, exec_{vm})$ for $vm \in VM'$ \label{al:assign:sort}
				\State $VM_0 \gets \argmin_{vm \in VM'}{trans_{t, vm}}$							\label{al:assign:select_near}
				\State $vm \gets VM_0[0]$														\label{al:assign:select}
				\State $T_{vm} \gets T_{vm} \cup \{t\}$											\label{al:assign:assign}
			\EndFor
			\State $P_{nl} \gets \{T_{vm}$ for $vm \in VM'\}$
			\State \Return $P_{nl}$
		\EndFunction
	\end{algorithmic}
\end{algorithm}

Algorithm \ref{al:assign} aims to evenly distributed tasks from $T'$ to the set of receiving VMs.

First of all, tasks are sorted in descending order based on their execution times (Line \ref{al:assign:tasks_sort}). Then, for each task, all the VMs which can execute it without requiring more than one time block is selected (Line \ref{al:assign:filter}). If there is no VM selected, i.e. it will take more than one time block if a task is assigned to any given VMs, the function fails (Lines \ref{al:assign:check} and \ref{al:assign:fail}).

All the selected VMs are sorted based on the distance between VM's location and the task's location, and by their current execution time (Lines \ref{al:assign:sort}). The task is assigned to the first VM in the sorted collection (Lines \ref{al:assign:select} and \ref{al:assign:assign}). In other words, Algorithm \ref{al:assign} tries to assign a task to the nearest VM with the lowest execution time.

\subsection{Reduce the Number of VMs}

\begin{algorithm}
	\caption{Reduce VMs}
	\label{al:reduce}
	\begin{algorithmic}[1]
		\Function{REDUCE}{$P, Ign, is\_local$}					\label{al:reduce:func}
			\State $vm \gets \argmin_{vm \in VM_{P}}{exec_{vm}}$		\label{al:reduce:select}
			\If{$is\_local = TRUE$}
				\State $VM' \gets VM_{l_{vm}} - vm$						\label{al:reduce:local}
			\Else
				\State $VM' \gets VM_P - vm$							\label{al:reduce:global}
			\EndIf
			\State $P' \gets ASSIGN(T_{vm}, VM')$						\label{al:reduce:reassign}
			\If{$tb_{P'} < tb_P$}										\label{al:reduce:check_tb}
				\State $P \gets P'$										\label{al:reduce:update}
			\Else
				\State $Ign \gets Ign \cup \{vm\}$						\label{al:reduce:ignore}
			\EndIf
			\If{$tb_P = tb_b$ or $\overline{Ign} = \overline{VM_P}$}	\label{al:reduce:constraint}
				\State \Return $VM_l$ for $l \in L$						\label{al:reduce:return}
			\Else
				\State \Return $LOCAL\_REDUCE(P_n, Ign)$				\label{al:reduce:continue}
			\EndIf
		\EndFunction
	\end{algorithmic}
\end{algorithm}


Algorithm \ref{al:reduce} is used to reduce the number of VMs by moving all tasks from one VM to others which are either in the same or on different locations. It is a recursive process which takes the current plan $P_n$, and the list of VMs which cannot be removed from the plan $Ign$, and the boolean value indicating if the reducing process is applied locally or globally $is\_local$.

First, a VM with lowest execution time is selected (Line \ref{al:reduce:select}). Then the remaining VMs, which can be either in the same (Line \ref{al:reduce:local}) or on different Cloud location (Line \ref{al:reduce:global}), are selected as receiving VMs.

After that, all tasks from selected VM are reassigned to other VMs (Line \ref{al:reduce:reassign}) by reusing the Algorithm \ref{al:assign}. Notably, the receiving VMs are not empty but already contain some tasks.

If the reassignment reduces the number of VMs (Line \ref{al:reduce:check_tb}), the current plan is updated (Line \ref{al:reduce:update}). Otherwise, the selected VM is added into the ignore list $Ign$ (Line \ref{al:reduce:ignore}). If the total time block satisfies the given constraint \textbf{or} all VMs are ignored (Line \ref{al:reduce:constraint}), the process stops and returns the current plan (Line \ref{al:reduce:constraint}), otherwise it continues (Line \ref{al:reduce:continue}).


\subsection{Balance Tasks Between VMs}

\begin{algorithm}
	\caption{Balancing Algorithm}
	\label{al:balance}
	\begin{algorithmic}[1]
		\Function{BALANCE}{$P$}
			\State $vm \gets \argmin_{vm \in VM_P}{exec_{vm}}$								\label{al:balance:select_vm}
			\State $T'_{vm} \gets T_{vm}$ sorted by $-exec_{t, vm}$							\label{al:balance:sort_tasks}
			\For{$t \in T'_{vm}$}
				\State $VM_1 \gets (VM_p - \{vm\})$ sorted by $trans_{t, vm}$
				\State $vm_0 \gets NULL$
				\For{$vm_1 \in VM_1$}
					\If{$t$ is never in $vm_1$}	AND {$rt{c_1} + exec_{t, c_1} < rt{c_0}$}	\label{al:balance:check}
						\State $vm_0 \gets vm_1$
						\State $BREAK$
					\EndIf
				\EndFor
				\If{$vm_0 \neq NULL$}
					\State $BREAK$
				\EndIf
			\EndFor
			\If{$vm_0 \neq NULL$}
				\State $T'_{vm} \gets T_{vm} - t$
				\State $T'_{vm_0} \gets T_{vm_0} \cup \{t\}$
				\State $P \gets (P - \{T_{vm}, T_{vm_0}\}) \cup \{T'_{vm}, T'_{vm_0}\}$
				\State \Goto{al:balance:select_vm}
			\EndIf
			\State \Return $P$
		\EndFunction
	\end{algorithmic}
\end{algorithm}

After the budget constraint is satisfied, the execution times between VMs can be uneven, i.e. some VMs can have higher execution times than the others. As the execution time of the plan $exec_P$ is based on the VM with highest execution time, it is necessary to balance out execution time between them.

Algorithm \ref{al:balance} is an iterative process which tries to move tasks from a VM with highest execution time (Line \ref{al:balance:select_vm}) to the nearest VM possible. There are two conditions for selecting a receiving VM: the selected task is never assigned to it and its execution time after receiving the task is not higher than the current execution time of the giving VM (Line \ref{al:balance:check}).

\subsection{Dynamic Scheduling To Avoid Idle VM}

\begin{algorithm}
	\caption{Dynamic Reassignment}
	\label{al:reassign}
	\begin{algorithmic}[1]
		\Function{REASSIGN}{$vm$}
			\If{$3600 - rt_{vm} < terminate\_time$}	\label{al:reassign:check}
				\State $FAIL$
			\EndIf
			\State $VM_1 \gets \{VM_P - \{vm\}\}$ sorted by $-e_{vm_1}$ for $vm_1 \in VM_1$	\label{al:reassign:sort}
			\State $vm_0 \gets NULL$
			\For{$vm_1 \in VM_1$}
				\If{$e_{vm_1} \leq thr_1$ AND $\overline{T_{r_{vm_1}}} \leq thr_2$}
					\State $vm_0 \gets vm_1$
					\State $BREAK$
				\EndIf
			\EndFor
			\If{$vm_0 = NULL$}
				\State $FAIL$
			\EndIf	\label{al:reassign:select:end_if}

			\State $T'_r \gets T_{r_{vm}}$ sorted by $trans_{t, vm}$ for $t \in T'_r$	\label{al:reassign:start_moving}
			\State $T \gets \emptyset$
			\State $el \gets 3600 - rt_{vm} - terminate\_time$
			\For{$t \in T'_r$}
				\State $exec'_{T} \gets exec_{T} + exec_{t, vm}$
				\If{$exec'_{T} \geq \frac{e_{vm_0} - thr_1}{2}$ OR $exec'_{T} > el$}
					\State $BREAK$
				\EndIf
				\State $T \gets T \cup \{t\}$
				\State $T'_r \gets T'_r - \{t\}$
			\EndFor	\label{al:reassign:end_moving}

			\State $T_{r_{vm}} \gets T_{r_{vm}} - T'_r$	\label{al:reassign:remove}
			\State $T_vm \gets T$	\label{al:reassign:add}

			\State $TIME\_OUT(vm, el)$ \label{al:reassign:timeout}
		\EndFunction
	\end{algorithmic}
\end{algorithm}

Even though Algorithm \ref{al:find} aims to build the plan in which all VMs finish their execution nearly at the same time, due to the instability of the network and other unaccountable factors, e.g. service failure, it is not unusual for one VM to finish before others. As the cost of a full hour is already paid, it is necessary to utilise the remaining time of the finished VMs in order to reduce not only idle and unpaid time but also the execution time of other VMs.

Let $rt_{vm}$ be the actual running time of a VM. Let $e_{vm}$ and $T_{r_{vm}}$ be the estimated remaining execution time and remaining tasks of $vm \in VM$. $terminate\_time$ denote the time it take for a VM to be shut down. Finally, $thr_1$ and $thr_2$ are 2 threshold values indicating the required remaining execution time and required remaining number of tasks. As unfinished VMs are still running when the reassignment is being performed, those thresholds aim to avoid reassigning tasks already executed by one VM to another. The idea of dynamic rescheduling is to move $T_{r_{vm}}$ of a VM to another finished one while satisfying $thr_1$ and $thr_2$ in order to reducing its $e_{vm}$.

In order to support dynamic scheduling, we add a feature which monitors the execution of VMs, keeps track of the remaining tasks and execution times, and detects a VM which has just finished its execution.

Every time there is a VM that has just finished its execution, Algorithm \ref{al:reassign} is invoked. First, it check whether there is enough time in a finished VM to execute some tasks (Line \ref{al:reassign:check}). This check ensures that the finished VM is able to be terminated before using another time block. Then, the VM which not only has the highest remaining execution time but also satisfies $thr_1$ and $thr_2$ is selected (Lines \ref{al:reassign:sort} to \ref{al:reassign:select:end_if}).

After that, some of the tasks are moved from the selected VM to the finished one until some conditions are met: i) the execution time of the finishes VM is greater or equal half of the remaining execution time of the giving one, or, ii) the finished VM will take more than one time block to finish its execution if more tasks are added (from Lines \ref{al:reassign:start_moving} to \ref{al:reassign:end_moving}).

Notably, Algorithm \ref{al:reassign} is only be invoked one at a time, i.e. if there are multiple finished VMs, only one of them is reassigned tasks while other have to wait.

Finally, the timeout feature is added to prevent the finished VM, which is just assigned some more tasks, to use more than one time block. Basically, it takes the VM and the allowed execution time as arguments (Line \ref{al:reassign:timeout}), if the VM is still running when time out, it is automatically terminated and the remaining tasks are moved to another VM with lowest remaining execution time, i.e. the one that is likely to finish first.


\section{\uppercase{Experimental Evaluation}}
\subsection{Set-up}

In order to evaluate our proposed approach, we developed a Word Count application in which each task involved downloading and counting the number of words in a file from a remote server. Those files were located on PlanetLab (PL), a test-bed for distributed computing experiments \cite{Chun:2003:POT}. We had 5700 tasks, i.e. files, distributed across 38 PL nodes and the total amount of data for each experiment run was more than 12 gigabytes. The VMs were deployed on 8 different Amazon Web Service (AWS) regions.


Prior to the experiment, we ran the test with fewer tasks in order to collect the computational cost, i.e. $comp$, and communicational costs between all AWS regions and PlanetLab Nodes (i.e. $trans$).

Based on our algorithm, at least 4 VMs were required to execute all 5700 tasks. We then set $tb_b = \{4, 6, 8, 10, 12, 14, 16, 18, 20\}$, i.e. the number of time block (or VMs) that we wanted to use. For each value of $tb_b$, we ran the execution three times to find the mean and standard deviation.

For comparison, we implemented two simple approaches for executing BoDT on the Cloud:

\begin{itemize}
	\item Centralised approach: one centralised location was selected as $l_c = \argmin_{l \in L}{(\sum_{t \in T}{trans_{t, l} * size_t})}$, i.e. the location whose the cost of moving all tasks to it was minimum in comparison with other locations. This approach was developed based on the centralised approach introduced in our previous paper \cite{Thai:2014:CloudCom}, however, instead of using only one VM at the selected location, in this paper, the number of VMs was equal to the one used by our proposed approach. In other words, this centralised approach enjoyed the same execution parallelism as the proposed one.
	\item Round Robin approach: for this approach, all Cloud locations was sorted in ascending order based on their costs of moving all tasks to them. Which means the first Cloud location was the one selected by the centralised approach. After that, VMs were added to each location in circular order, e.g. the first VM was added to the first Cloud location in the sorted list.
\end{itemize}

For both approaches, Algorithm \ref{al:assign} was used to evenly distribute tasks to all VMs.

\subsection{Dynamic Reassignment}

\begin{figure}[h!]	\centering
		\includegraphics[width=0.5\textwidth, height=0.25\textheight]{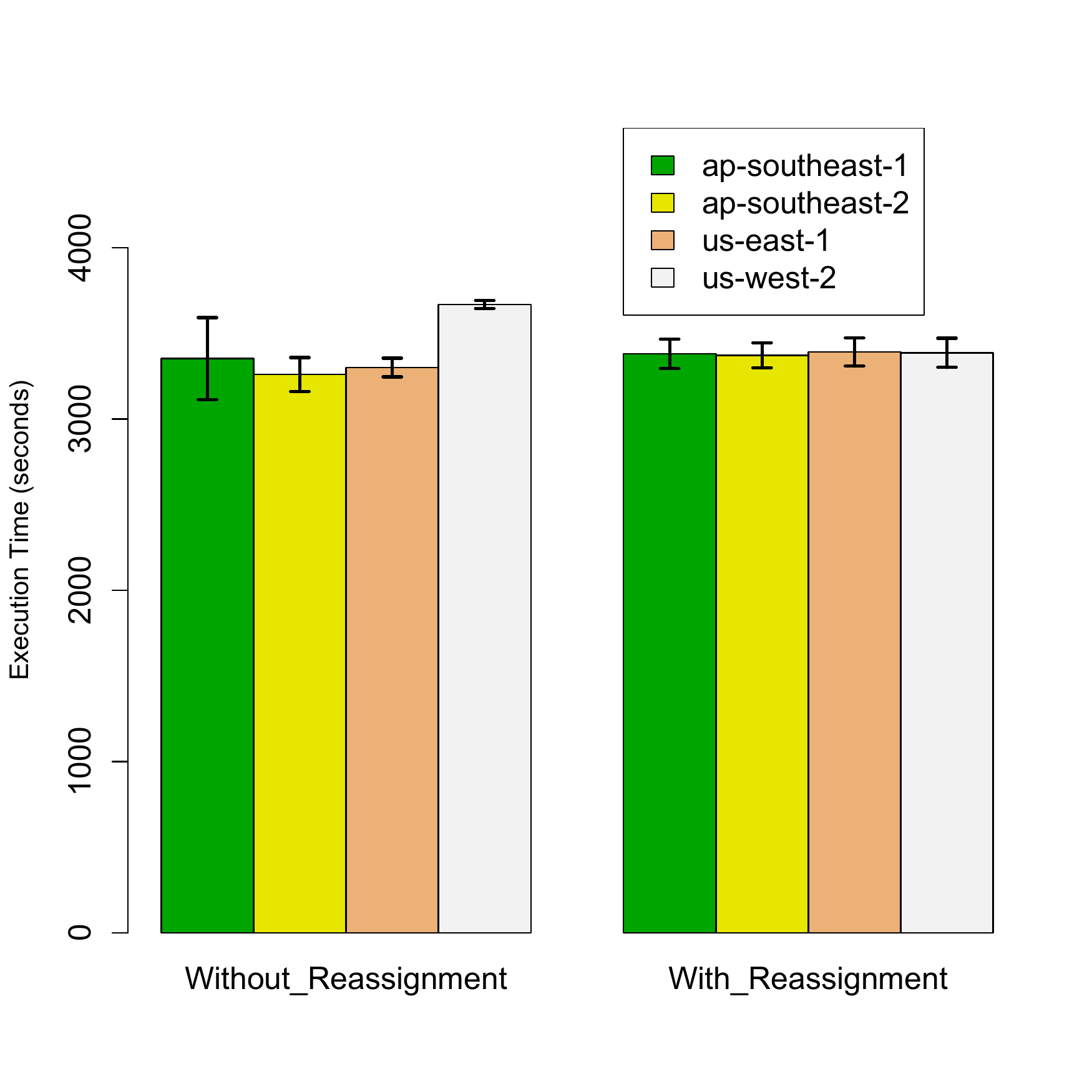}
	\caption{Compare execution without and with reassignment}
	\label{fi:reassignment}
\end{figure}

Before going into the main experiment, it is necessary to demonstrate the need of using dynamic reassignment for VMs that finish executing their assigned tasks earlier than others. Figure \ref{fi:reassignment} presents the result of running the same execution plan with $tb_b = 4$, i.e. there were 4 VMs. Each bar represents the execution time of a VM. Without reassignment, one VM took longer to finish its execution thus increasing the overall execution time. Dynamic reassignment helped to balance out the execution time between VMs so that all VMs could finish at about the same time, which in turn reduced the overall execution time.

For the remainder of the experiments presented in this section, dynamic reassignment is applied.

\subsection{Experimental Results}

\begin{figure}[h!]	\centering
		\includegraphics[width=0.5\textwidth]{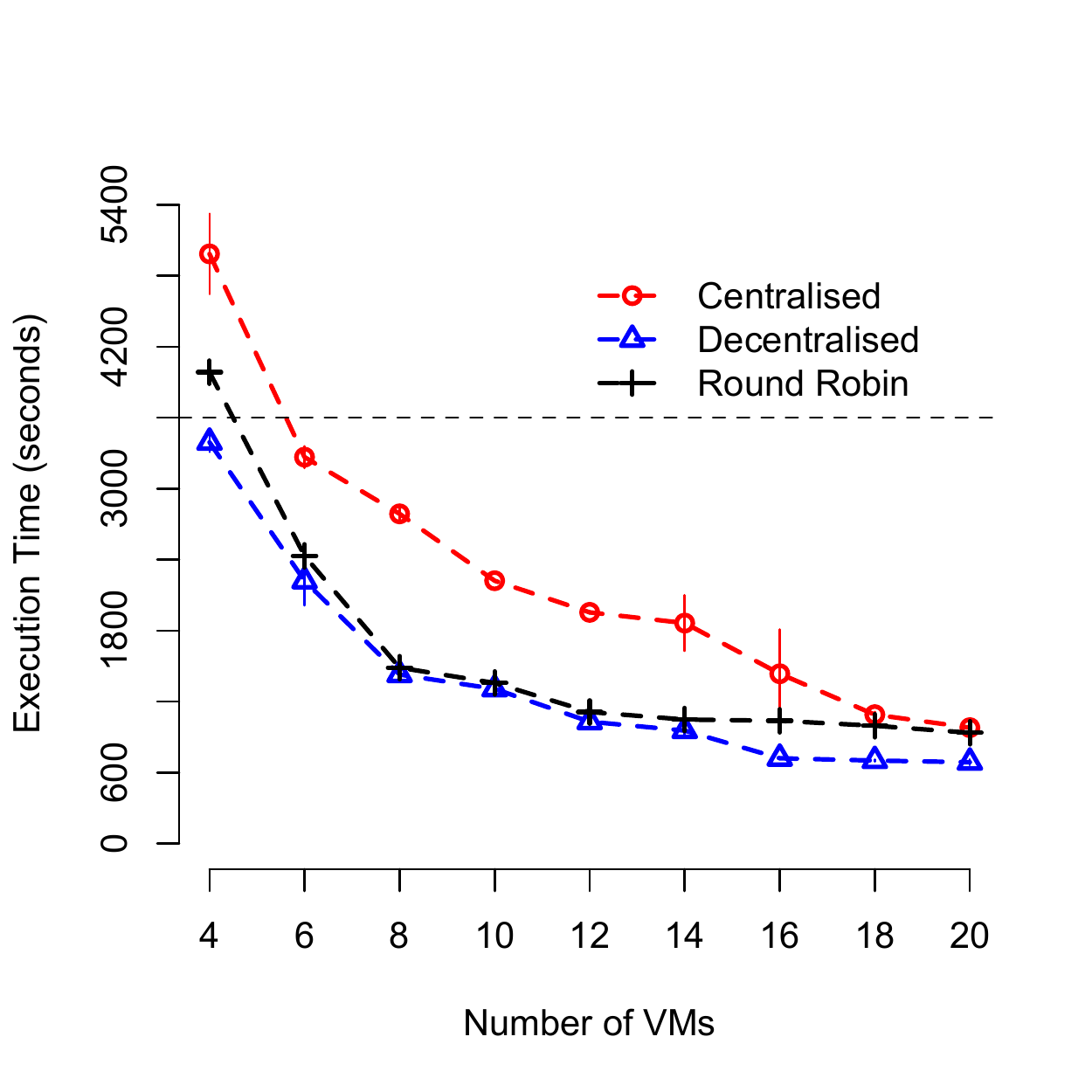}
	\caption{Execution Times}
	\label{fi:comp_exec_times}
\end{figure}

Figure \ref{fi:comp_exec_times} presents the execution times corresponding for each value of the number of VMs for all three approaches. The centralised approach had the highest execution times as even though it selected the location with lowest transfer cost for all tasks but some tasks were very far from the Cloud location which resulted in the high data transfer time. On the other hand, the round robin approach performed better as it deployed VMs at multiple Cloud locations, which means it was possible for tasks to be executed near their data sources. Finally, it is evidently to see that, with the same number of VMs (or budget), our approach always had the lowest execution time, i.e. performed better, in comparison with other two. 

A reason for the improvement is that our approach not only deployed VMs at multiple locations but also carefully selected those locations so that the majority of tasks could be executed near their data sources. The two simple approaches decided the location(s) of VMs based on \textbf{all} tasks, by assuming all tasks were assigned to one Cloud location. On the other hand, our approach took a more fine-grain method by assigning each task to its nearest location first and then reassigning them to others location until the budget constraint was satisfied.

As the result, with the same given budget constraint, our approach was 30\% to 50\% faster than the centralised approach. In comparison to the round robin approach, ours was able to reduce the execution times up to 30\%.

\begin{figure}[h!]	\centering
		\includegraphics[width=0.5\textwidth, height=0.25\textheight]{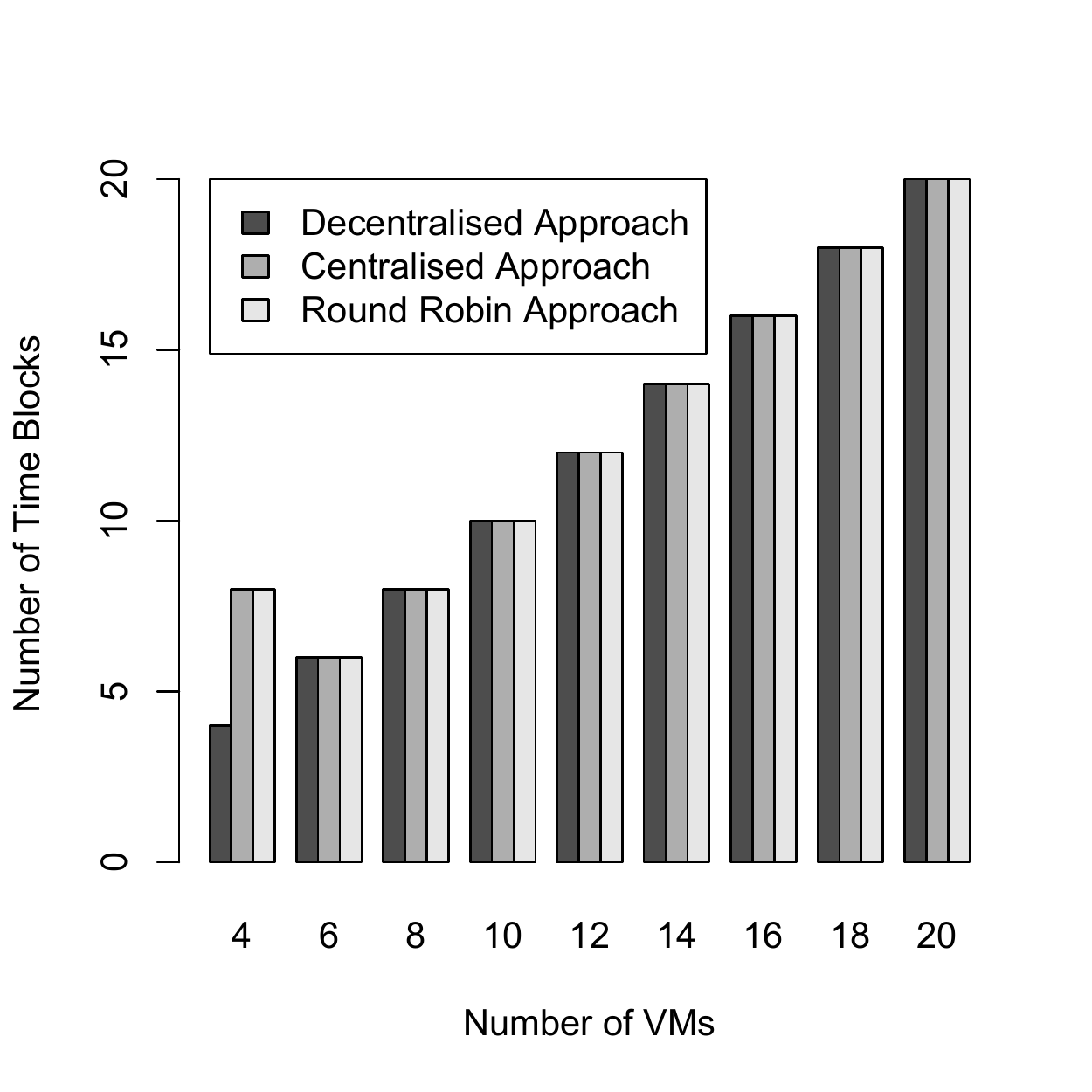}
	\caption{Actual Number of Used Time Blocks, i.e. cost}
	\label{fi:budgets}
\end{figure}

Figure \ref{fi:budgets} presents the number of actual time blocks, which can be mapped onto actual cost, consumed by three approaches. It shows that our approach was able to satisfy the budget constraint in all cases. Moreover, when there were 4 VMs, the centralised and round robin approaches were more expensive than the decentralised one. It was because each of their VMs required more than one hour to finish executing all the assigned tasks and the overall execution time was higher than 3600 seconds, as shown by Figure \ref{fi:comp_exec_times}. Which means that the constraint $tb_b = 4$ could only be satisfied by the decentralised approach.

\subsection{Trade-off Between Cost and Performance}

As presented in Figure \ref{fi:comp_exec_times}, the higher the budget constraint is (i.e. more VMs), the better the performance is. In theory, it is possible to keep adding more VMs in order to achieve better performance. However, the performance gain for each additional VM also decreases as the total number of VMs increases.

Hence, it is up the user to decide how much improvement in performance can be afforded. There are some simple criteria to consider such as a defined budget constraint, the desired execution time or defining a threshold in the performance gain (for example, stop adding more VM(s) if the performance gain is less than 60 seconds).

A user can also make the decision of how many VMs to use based on the trade-off between performance and cost, as mentioned in \cite{Thai:2014:CloudCom}.

\section{\uppercase{Related Work}}
In Grid environment, in which the resources are shared between multiple organisations, \cite{Ranganathan:2002:DCD} was able to improved the overall performance of a distributed framework by processing data in close proximity to where it resided. Similarly, the authors of \cite{Kaya:TransPDS:2006} proposed a heuristic algorithm to improve performance in executing independent but file-sharing tasks. In \cite{Venugopal:2005:ICA3PP}, the authors assumed that each task required data which was distributed at multiple sources and proposed the auto-scaling algorithm to satisfy both deadline and budget constraints.

However, the application of Grid computing research on Cloud computing is limited because: i) the Cloud resources are (virtually) unlimited, hence a user is free to add or remove VMs whenever she wants but ii) the monetary cost factor has to be considered as the resource is no longer free-of-charge.


Recently, running application on the Cloud has received attention from many researchers. Statistical learning had been used to schedule the execution of BoT on the Cloud \cite{Oprescu:2010:CloudCom}. The method for scaling resource based on given budget constraint and desired application performance was also investigated \cite{Mao:2010:GRID}. Nevertheless, those papers did not consider the location of data.

Cloud computing is employed for improving the performance of data intensive application, such as Hadoop, whose data is globally located \cite{Ryden:2012:IC2E}. Research that takes geographical distance into account while executing workflows is reported in \cite{Luckeneder:2013:CloudCom,Thai:2014:CloudCom:WF}. However, recent researches on applying Cloud computing for applications with geographically distributed data only focus on improving the performance without considering the monetary cost.

Our previous work \cite{Thai:2014:CloudCom} aimed to determine a plan for executing BoDT on the Cloud, however, it made an assumption that there was only one VM that could be deployed at each Cloud region.


Our paper differentiates itself from prior research by taking advantage of the decentralised infrastructure of Cloud computing in executing BoDT application. We tries to decide not only the amount of resources but also the locations where resources, i.e. VMs, must be located. Moreover, our research exploits of the virtually unlimited resources of Cloud computing by letting a user decides how much resources that she wants based on her budget. Finally, the trade-off between performance gain and additional cost is also presented.

\section{\uppercase{Conclusion}}
Due to its decentralised infrastructure and virtually unlimited resources, Cloud computing is suitable to execute BoDT, whose data is globally distributed all over the world. It is challenging to decide how to assign tasks to Cloud VMs based on a user's budget constraint while minimising the execution time. 

The above problem was mathematically modelled in this paper. We also proposed a heuristic approach which assigned BoDT to Cloud VM(s) in order to maximise performance and to satisfy the allowed cost provided by a user.



Furthermore, we implemented a dynamic reassignment feature to utilise the idle time of a VM that completes execution ahead of others by assigning tasks from other VMs onto it. This feature reduces the overall execution time when a number of VMs take longer to finish their execution due to service failure or network instability.

Our approach was evaluated and able to provide execution plans which satisfied given budget constraints. Compared to the centralised and round robin approaches, our approach reduced the execution time by average 27\%. Our approach was also able to satisfy the low budget while the others could not.



In the future, we plan to further improve dynamic resource provisioning and tasks scheduling so that they can be performed during execution in order to handle expected events, e.g. network instability or machine failure. Moreover, the different types of Cloud instances, which have varying performance and cost, will be taken into account. 

\section*{\uppercase{Acknowledgements}}

\noindent This research is supported by the EPSRC grant `Working Together: Constraint Programming and Cloud Computing' (EP/K015745/1), a Royal Society Industry Fellowship, an Impact Acceleration Account Grant (IAA) and an Amazon Web Services (AWS) Education Research Grant.

\vfill
\bibliographystyle{apalike}
{\small
\bibliography{ref}}

\begin{thebibliography}{}

\bibitem[Chun et~al., 2003]{Chun:2003:POT}
Chun, B., Culler, D., Roscoe, T., Bavier, A., Peterson, L., Wawrzoniak, M., and
  Bowman, M. (2003).
\newblock Planetlab: An overlay testbed for broad-coverage services.
\newblock {\em SIGCOMM Comput. Commun. Rev.}, 33(3):3--12.

\bibitem[Kaya and Aykanat, 2006]{Kaya:TransPDS:2006}
Kaya, K. and Aykanat, C. (2006).
\newblock Iterative-improvement-based heuristics for adaptive scheduling of
  tasks sharing files on heterogeneous master-slave environments.
\newblock {\em Parallel and Distributed Systems, IEEE Transactions on},
  17(8):883--896.

\bibitem[Luckeneder and Barker, 2013]{Luckeneder:2013:CloudCom}
Luckeneder, M. and Barker, A. (2013).
\newblock Location, location, location: Data-intensive distributed computing in
  the cloud.
\newblock In {\em In Proceedings of IEEE CloudCom 2013}, pages 647--653.

\bibitem[Mao et~al., 2010]{Mao:2010:GRID}
Mao, M., Li, J., and Humphrey, M. (2010).
\newblock Cloud auto-scaling with deadline and budget constraints.
\newblock In {\em Grid Computing (GRID), 2010 11th IEEE/ACM International
  Conference on}, pages 41--48.

\bibitem[Oprescu and Kielmann, 2010]{Oprescu:2010:CloudCom}
Oprescu, A. and Kielmann, T. (2010).
\newblock Bag-of-tasks scheduling under budget constraints.
\newblock In {\em Cloud Computing Technology and Science (CloudCom), 2010 IEEE
  Second International Conference on}, pages 351--359.

\bibitem[Ranganathan and Foster, 2002]{Ranganathan:2002:DCD}
Ranganathan, K. and Foster, I. (2002).
\newblock Decoupling computation and data scheduling in distributed
  data-intensive applications.
\newblock In {\em Proceedings of the 11th IEEE International Symposium on High
  Performance Distributed Computing}, HPDC '02, pages 352--, Washington, DC,
  USA. IEEE Computer Society.

\bibitem[Ryden et~al., 2014]{Ryden:2012:IC2E}
Ryden, M., Oh, K., Chandra, A., and Weissman, J.~B. (2014).
\newblock Nebula: Distributed edge cloud for data intensive computing.

\bibitem[Thai et~al., 2014a]{Thai:2014:CloudCom:WF}
Thai, L., Barker, A., Varghese, B., Akgun, O., and Miguel, I. (2014a).
\newblock Optimal deployment of geographically distributed workflow engines on
  the cloud.
\newblock In {\em 6th IEEE International Conference on Cloud Computing
  Technology and Science (CloudCom 2014)}.

\bibitem[Thai et~al., 2014b]{Thai:2014:CloudCom}
Thai, L., Varghese, B., and Barker, A. (2014b).
\newblock Executing bag of distributed tasks on the cloud: Investigating the
  trade-offs between performance and cost.
\newblock In {\em Cloud Computing Technology and Science (CloudCom), 2014 IEEE
  6th International Conference on}, pages 400--407.

\bibitem[Venugopal and Buyya, 2005]{Venugopal:2005:ICA3PP}
Venugopal, S. and Buyya, R. (2005).
\newblock A deadline and budget constrained scheduling algorithm for escience
  applications on data grids.
\newblock In {\em in Proc. of 6th International Conference on Algorithms and
  Architectures for Parallel Processing (ICA3PP-2005}, pages 60--72.
  Springer-Verlag.

\end{thebibliography}

\end{document}